\documentclass[10pt,twocolumn]{article}
\usepackage[T1]{fontenc}
\usepackage{verbatim}
\usepackage{mathptmx}
\usepackage{listings}
\usepackage{courier}
\usepackage{geometry}
\usepackage{caption}
\pdfoutput=1
\usepackage[T1]{fontenc}
\usepackage{hyperref}
\usepackage{url}
\usepackage{dcolumn}
\usepackage{graphicx}
\usepackage{color}
\usepackage{geometry}

\usepackage{listings}
\usepackage{color}

\definecolor{mygreen}{rgb}{0,0.6,0}
\definecolor{mygray}{rgb}{0.5,0.5,0.5}
\definecolor{mymauve}{rgb}{0.58,0,0.82}

\begin{document}

\title{\vspace{-1em}\huge{\textbf{Warrior1: A Performance Sanitizer for C++}}}

\author{Nadav Rotem, Lee Howes, David Goldblatt \\ Facebook, Inc.}
\date{\today}

\maketitle

\section{Abstract}

This paper presents Warrior1, a tool that detects performance anti-patterns in
C++ libraries. Many programs are slowed down by many small inefficiencies.
Large-scale C++ applications are large, complex, and developed by large groups
of engineers over a long period of time, which makes the task of identifying
inefficiencies difficult.
Warrior1 was designed to detect the numerous small performance issues that
are the result of inefficient use of C++ libraries. The tool detects performance
anti-patterns such as map double-lookup, vector reallocation, short lived
objects, and lambda object capture by value.
Warrior1 is implemented as an instrumented C++ standard library and an off-line
diagnostics tool. The tool is very effective in detecting issues. We demonstrate
that the tool is able to find a wide range of performance anti-patterns in a
number of popular performance sensitive open source projects.

\section{Introduction}

One of the challenges of modern C++ is the ability to write
performance-predictable code using high-level abstractions.  The C++ Standard
Template Library (STL) provides data structures such as string, vector, map,
list and smart pointers. These data structures are frequently used in C++
programs. These high-level data structures abstract away the low-level details,
and in many cases hide performance issues.  It is often difficult to reason
about the performance of C++ programs without testing the code on real inputs.
It's hard to known when the code might allocate or copy large chunks of memory.
Consider the code below.

\lstset {language=C++}
\begin{lstlisting}
  std::vector<int> v;
  for (int i = 0; i < n; i++)
    v.push_back(i);
\end{lstlisting}

The standard C++ vector represents a sequence of elements that can grow
dynamically. The amortized growth cost of vector is constant. Every time the
vector underlying storage grows the standard library allocates a new buffer, copies the
data and deletes the old buffer. As the vector grows the buffer size expands in a geometric sequence.
Constructing a vector of 10 elements in a loop results in 5
calls to 'malloc' and 4 calls to 'free'. These operations are relatively
expensive. Moreover, the 4 different buffers pollute the cache and make the
program run slower.

One way to optimize the performance of this code is to call the 'reserve' method
of vector. This method will grow the underlying storage of the vector just once
and allow non-allocating growth of the vector up to the specified size.  The
vector growth reallocation is a well known problem, and there are many other
patterns of inefficiency, some of which are described in section \ref{rules}.

\lstset {language=C++}
\begin{lstlisting}
  std::vector<int> v;
  v.reserve(n);
  for (int i = 0; i < n; i++)
    v.push_back(i);
\end{lstlisting}

\subsection{The long tail of performance issues}

In many programs, a small number of issues have a significant impact on
the slowness of the program, but there are also many issues that slow down the
program each by a small amount. It is easy to identify and fix the egregiously
slow parts of the code.  However, it's often difficult to identify the long list
of program inefficiencies.  But these issues add up and make the system slow.

Software engineers often use profilers to identify hot spots, which are areas of
the code in which the program spends a lot of time. This method is very
effective in finding the major performance issues. However, profilers are not
good at detecting the long tail of issues. A long list of small issues produce a
flat profile, where the time is spent in many different places in the program.
Programmers refer to the long tail of issues as "Death by a thousand cuts".
Slowness is difficult to measure because one part of the code may
evict the cache, and another unrelated procedure in a different part of the code
may pay the price and run more slowly.  
Another issue is that programmers lack an a priori expectation for how fast a
function should be. If a program spends 0.3\% in some parts of the code, is that
a lot?

Warrior1 was designed to catch the long tail of inefficient operations that only
together add-up to a meaningful performance gain.
The Warrior1 tool generates detailed reports that point out the exact location
in the source code where the performance issues are located. The performance
report is sorted by the severity of the issue. Figure \ref{fig:report1} brings a
section of the report that was generated for the Redex project.

\begin{figure}
  \includegraphics[width=0.5\textwidth]{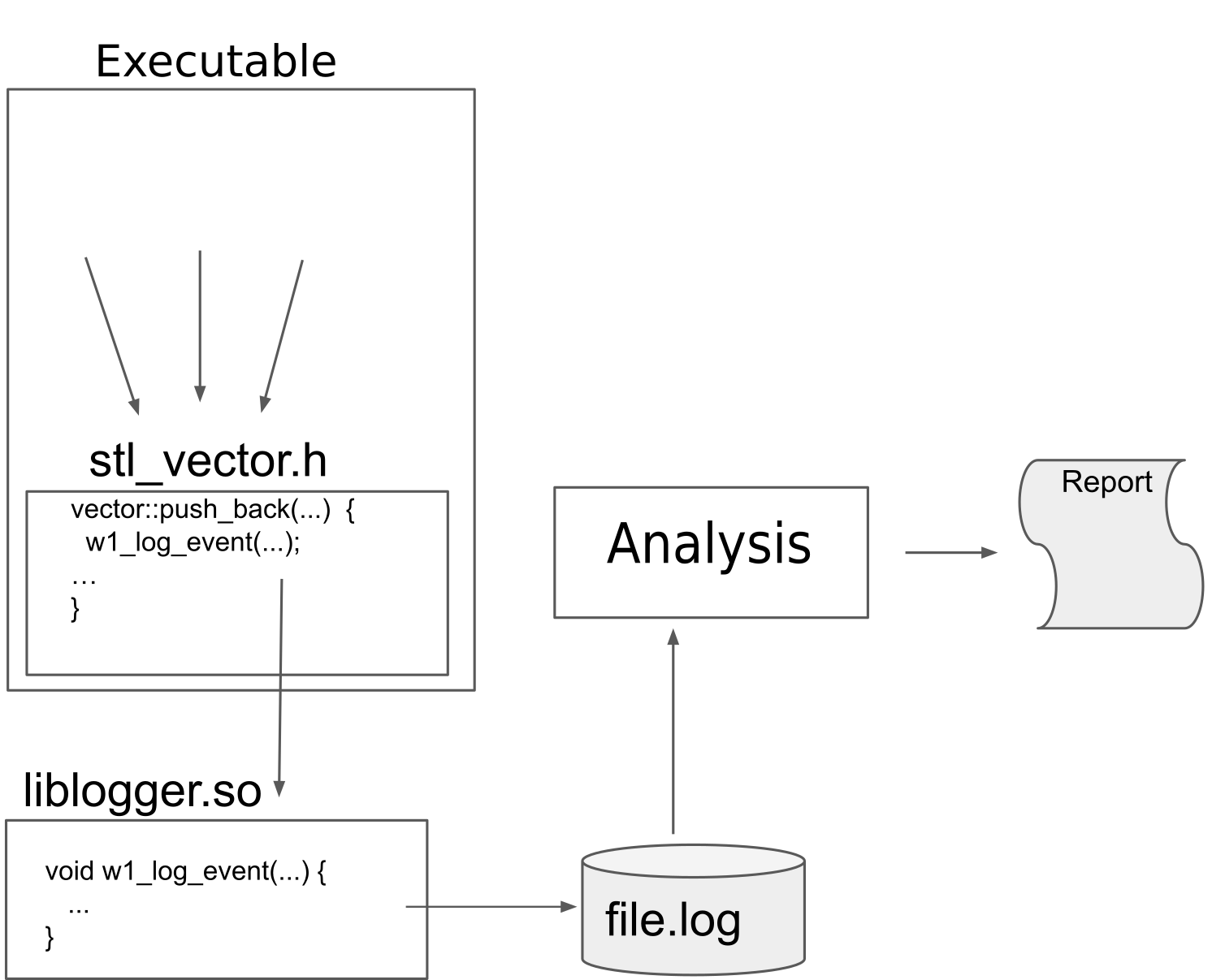}
  \caption{System overview}
  \label{fig:system}
\end{figure}

\begin{figure*}
\lstset {language={}}
\begin{lstlisting}[]
** Repeatedly growing a vector (total 177,723 reallocations) in 85,571 instances.
** Consider reserving space when the vector is constructed.

0) 0x9bbd4b void std::vector<cfg::Block*, std::allocator<cfg::Block*> >::_M_realloc_insert<
        /toolchain/include/c++/10.2.1/bits/vector.tcc:441
1) 0x9fbbdd std::_Function_handler<std::vector<cfg::Block*, std::allocator<cfg::Block*>>
        /redex/redex/sparta/include/MonotonicFixpointIterator.h:530
        /toolchain/include/c++/10.2.1/bits/std_function.h:292
2) 0x9c1dfd sparta::wpo_impl::WpoBuilder<cfg::Block*, std::hash<cfg::Block*> >::construct
        /toolchain/include/c++/10.2.1/bits/stl_vector.h:882
        /redex/redex/sparta/include/WeakPartialOrdering.h:357
3) 0x9c2f89 sparta::WeakPartialOrdering<cfg::Block*, std::hash<cfg::Block*> >::WeakPartial
        /redex/redex/sparta/include/WeakPartialOrdering.h:280
        /redex/redex/sparta/include/WeakPartialOrdering.h:194


** Map instance is not used as an ordered containers (in 472 instances).
** Consider using std::unordered_map instead.

0) 0x9ad460 IRTypeChecker::run()
        /redex/redex/libredex/IRTypeChecker.cpp:393
        /redex/redex/libredex/IRTypeChecker.cpp:565
1) 0xa0e67f CheckerConfig::run_verifier(std::vector<DexClass*,
        /redex/redex/libredex/IRTypeChecker.h:112
        /redex/redex/libredex/PassManager.cpp:186
2) 0xa0f305 std::thread::_State_impl<std::thread::_Invoker<std::
        /redex/redex/libredex/Trace.h:174
        /redex/redex/libredex/Walkers.h:439
        /redex/redex/libredex/WorkQueue.h:24
        /toolchain/include/c++/10.2.1/thread:264
\end{lstlisting}
  \caption{Sample report for the Redex program}
  \label{fig:report1}
\end{figure*}

\subsection{How Warrior1 works}

Warrior1 is built like a logger that records some of the methods in the C++
standard library. A good analogy would be the description of adding a call to
'printf' inside every method that reallocates a vector, string or a map.
Analysis of the generated log, sorted by the instance pointer, can reveal
inefficiencies.  Any vector instance has enough information to know that it was
resized several times.  Of course that this is only an analogy and the actual
implementation that takes into consideration the accuracy of the report and the
performance of the system is more involved.

Warrior1 has several parts. First, a compiler toolchain that contains an
instrumented standard library that contains calls to one specific log function.
Every constructor and destructor and some methods contain a call to the log
function. The second part in the system is the logger. The logger is called from
different source files in the standard library. It records the current time, a
stack trace and other relevant pieces of information. The logger compresses the
events, coalesces them into large packets and writes them to disk. The third
piece is a reporting tool that loads the log files, analyzes them and reports a
list of warnings and suggestions.

\section{Implementation}

This section describes the implementation of the system.
Figure \ref{fig:instance} depicts a log of a single vector instance. The
numbers that follow the method names are the parameters that are passed to the
logger.  The logger records the size and capacity of std::vector. The last
number in the listing is the stack-trace id. The timestamp is omitted from the
example.  It's easy to figure out from the log that the user calls the push\_back
API, and increases the size of the vector and that the vector re-allocates
several times in the calls to the 'slow-path' method.

In order to generate efficient diagnostics the logger needs to records many
different pieces of information: A full stack trace that reports the caller
information. Stack traces typically have dozens of entries, each one is a 64-bit
pointer. The logger also needs to record a 64-bit timestamp for each method.
The logger records a 64-bit 'this' pointer that points to the address of the
instance.  The logger records the name of the class (such as "vector") and a
string that contains the name of the method (such as "push\_back"). The logger
also records three 32-bit integers with additional information about the
instance. For example, the std::string destructor reports the size, capacity and
hash of the string content.  In addition to the per-method information the
logger records the names and addresses of all of the loaded shared objects in
memory to allow off-line symbolication.

\subsection{Log Compression}

The methods in the C++ standard library are invoked millions of times in a short
period of time. A naive implementation that writes uncompressed logs on each and
evey event would be very slow and generate huge log files. The Warrior1 logger
implements several optimizations that make the approach practical. On average
method events are compressed and encoded as three 64-bit words using techniques
as side tables and path compression.

When writing data into a file, the program has to make several system calls to
open the file, append data at the end, and close the file. System calls and file
IO are very slow and the system needs to minimize the number of operations.
Warrior1 reduces the number of file-system operations by writing into a local
memory buffer of several megabytes. The buffer is flushed to disk when enough
data was accumulated.

The logger maintains a protocol that sends commands that the analysis tool can
interpret.  The commands help to compress the log file and the messages that are
sent. For example, one of the basic commands that the logger sends is the
command 'register string', that contains a string literal and a key integer
value.  Following packets can send textual information by using the integer that
was previously specified. This allows the logger to send the string
'shared\_ptr' only once in a session, and use a 16-bit integer to refer to this
string in every log entry.

The logger sends the following commands:
\begin{itemize}
 \item Register a new string.
 \item Register a new stack trace.
 \item Register a new shared-object section.
 \item A new stack-trace event.
 \item A new compact stack-trace event.
\end{itemize}

Stack traces are a list of pointers that point to executable code and describe
the call-stack, which is the execution context of the current function. Every
entry in the call stack is a 64-bit pointer, and there are usually dozens of
entries in the stack trace. The main function is usually one of the first
entries in the stack trace. Saving the full call stack is important for
providing readable location information to the user, but saving around 16 64-bit
numbers on each method invocation is unrealistic because so much data is
written. Instead Warrior1 takes a different approach to saving stack traces. The
logger maintains a Trie data structure. Stack traces are inserted into the Trie, where each node represents a frame address.
Each leaf node is assigned with a unique 32-bit id. The logger sends the 
integer that identifies the stack trace. The logger also sends a new-node
registration command on the first occurrence, but these commands are sent at a
much lower frequency.

The logger can send two kinds of event commands: regular and compact. The
regular event command lists the relevant log fields: timestamp, stack-trace id,
this pointer, etc. In some cases the logger is able to fit all of the fields
into a compact command. This happens when the parameters to the logger are small
values that fit in fewer bits. The STL methods often log information such as the
number of elements in the container, and zero and other small values can be
encoded efficiently.  Over 90\% of the event commands are compact event commands
that are encoded in only 3 64-bit numbers. These 3 64-bit numbers represent a
complete entry that describes an event such as "a shared\_ptr object of address
X, with stack trace Y was destroyed and the reference count was 1.".

\begin{figure}
\lstset {language=Bash}
\begin{lstlisting}[]
Instance 0x7ffefb013b20:
vector::vector      [3, 0, 0]   loc[0xe98] 
vector::push_back   [3, 3, 0]   loc[0xf4e]
vector::__slow_path [0, 0, 0]   loc[0x096]
vector::push_back   [4, 6, 0]   loc[0x0a3]
vector::push_back   [5, 6, 0]   loc[0x0ab]
vector::push_back   [6, 6, 0]   loc[0x0ab]
vector::__slow_path [0, 0, 0]   loc[0x096]
vector::push_back   [7, 12, 0]  loc[0x0a3]
vector::push_back   [8, 12, 0]  loc[0x0ab]
vector::push_back   [9, 12, 0]  loc[0x0ab]
vector::push_back   [10, 12, 0] loc[0x0ab]
vector::push_back   [11, 12, 0] loc[0x0ab]
vector::push_back   [12, 12, 0] loc[0x0ab]
vector::__slow_path [0, 0, 0]   loc[0x096]
vector::push_back   [13, 24, 0] loc[0x0a3]
vector::push_back   [14, 24, 0] loc[0x0ab]
\end{lstlisting}
  \caption{A log of one vector instance.}
  \label{fig:instance}
\end{figure}

\subsection{Symbolication}

Symbolication is the procedure of assigning source locations to a stack trace.
The Warrior1 logger records the list of loaded ELF sections and the addresses of
their text sections. Addresses in the stack traces should fall within the
regions of memory that represent the code section of the loaded shared objects
(the actual name of the code section is "text").  The logger does not save the
actual shared objects, so Symbolication needs to happen on the same file-system
on which the recording was done.

There are two methods for symbolicating addresses. Non-stripped executables and
shared objects contain a list of symbols with their addresses. Tools such as
"addr2line" find the symbol closest to the searched address. They return strings
that represent the mangled name plus some byte offsets into the function. The
second method, which is more accurate, is the use of DWARF debug information.
The debug information in the binary records more information that includes
things like functions that were inlined and source code information. Warrior1
uses the LLVM\cite{llvm} infrastructure to symbolicate stack traces and present
accurate stack traces with detailed source information.

\subsection{Report Tool}

The report tool is an executable that loads the log file and prints diagnostic
reports. The tool is built as a simple processing pipeline with several stages.
The processing pipeline starts by loading the log file from disk and
reconstructing the state that was saved by the logger. The report tool builds
the stack trace Trie, expands strings and builds the memory map of remote ELF
sections. Next, the report tool sorts the class instances into separate
buckets. This is done by creating lists of command events that are indexed by
the address of the instance (the 'this' pointer). Notice that the memory
allocator may reallocate new instances on previously freed addresses. In order
to create an accurate picture, the log parser breaks instances into different
buckets by detecting destructors. The parser uses names like '~basic\_string' to
figure out that the instance has been destructed and any future method calls are
a part of a new instance that was allocated at the same address.  The output of
this phase creates data structures similar to the example in figure
\ref{fig:instance}.  Next, the tool detects performance issues for all of the
object instances in the program. And finally the tool sorts the warnings
according to their severity before printing them to the user.

\subsection{Diagnostics Rules}  \label{rules}
The report tools implements several rules that expose inefficient patterns:

\begin{itemize}

\item Short-lifetime: the report tool detects instances that are alive for a very short period of time. This rule exposes temporary objects and objects that are created as part of temporary expressions.

\item Vector and String growth-reallocation: this rule detects strings and vectors that grow multiple times and recommends to reserve space when the object is created.

\item Data shift: this rule detects cases where elements are inserted into the low parts of containers and many elements are shifted.

\item Push\_back of non-copyable elements: this rule detects cases where emplace\_back should be used instead of the push\_back method.

\item Shrink vectors to save memory: this rule detects cases where a vector can be shrunk in order to free the unused remaining capacity (by calling shrink\_to\_fit).

\item Object and Value copy detection: a set of rules that detect passing of objects by value, and copy of heap-allocated instances that contain object members.

\item A set of rules that detect vectors that remain small and can fit in small vectors.

\item Unique shared pointer: a rule that detects shared\_ptr that has a reference-count of 1, and can be converted to a unique\_ptr.

\item Duplicate string instance: this rule compares the hash of the string content and compares it to previously-hashed strings. This rule detects string contents that are saved many times in memory in different places.

\item Unnecessarily ordered map: this rule detects standard maps that are not traversed in a specific order. The recommendation is to convert the instance to unordered maps that are more efficient.

\item Double-map lookup: this rule detects the pattern of 'operator[]' after 'count', which is a common map anti-pattern.

\item Unused instance: this rule detects instances that are constructed and never used. We pay the cost of constructing the object but never use it.

\item Shared pointers with a very high reference count.

\end{itemize}

\subsection{Visualization}

The information that is collected during the execution of the program allows the
tool to visualize different aspects of the program. In this section we bring a
few reports from open source programs.

\begin{figure}
  \includegraphics[width=0.5\textwidth]{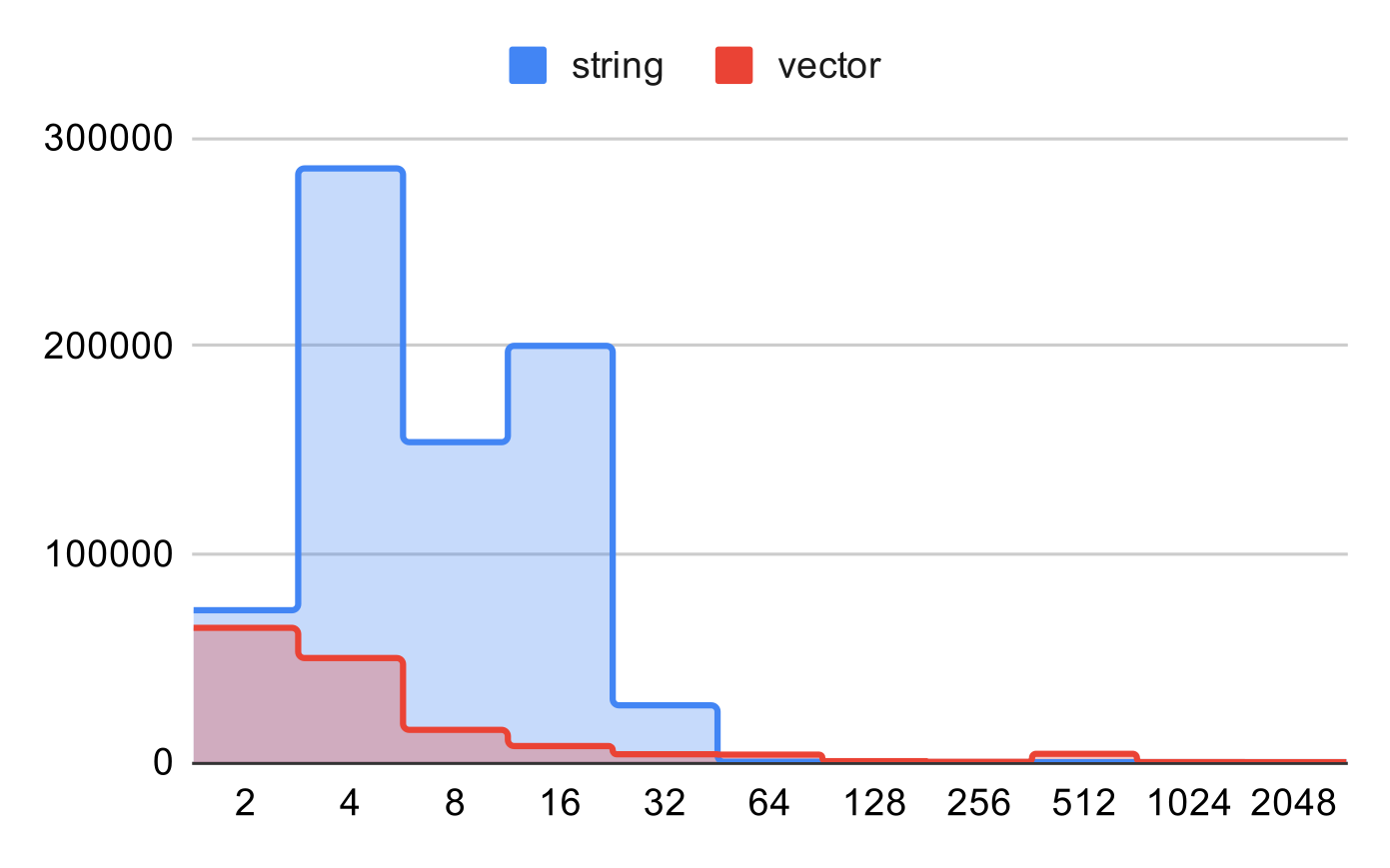}
  \caption{Number of instances in different size-buckets in Clang-11.}
  \label{fig:hist2}
\end{figure}

\begin{figure}
  \includegraphics[width=0.5\textwidth]{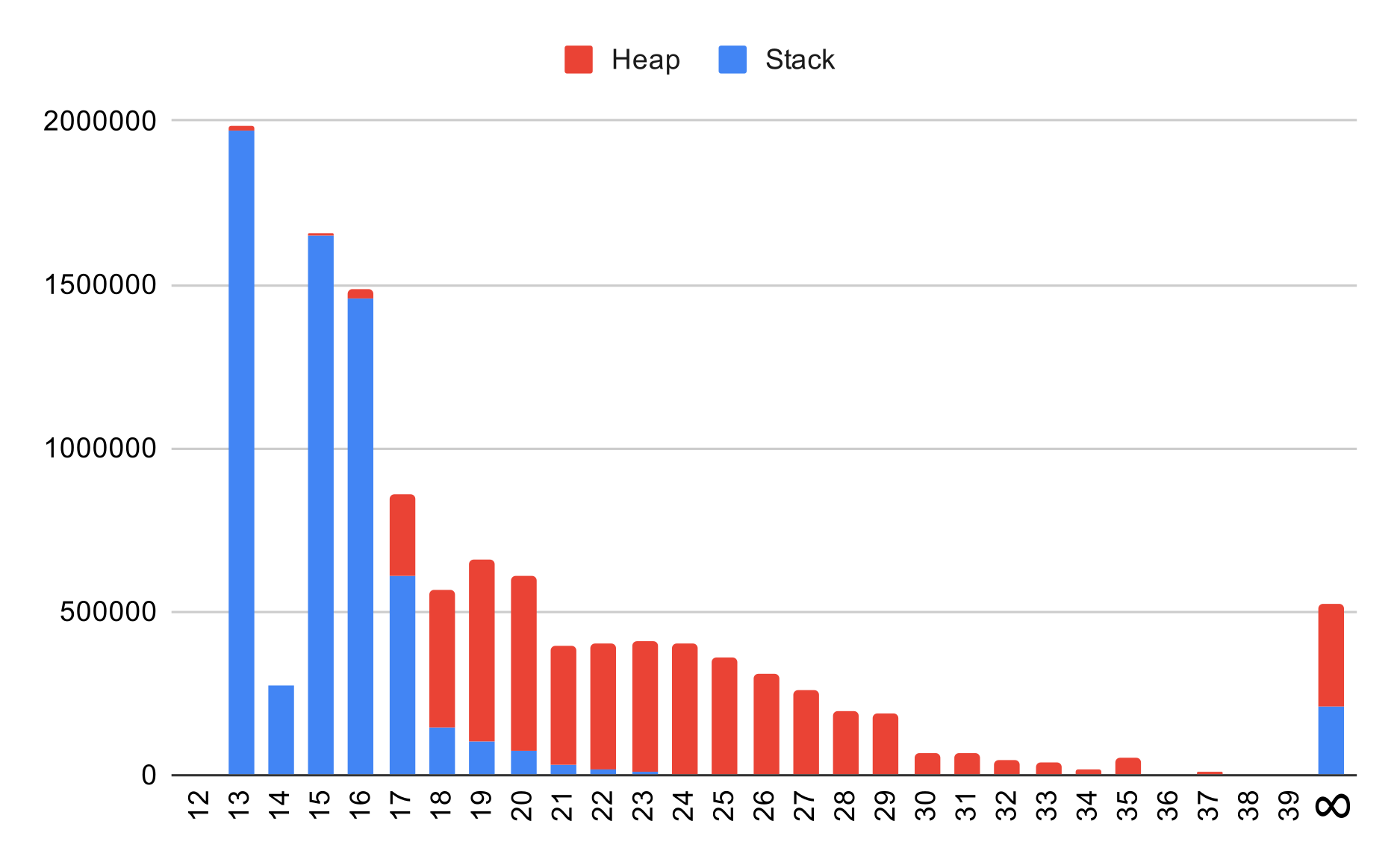}
  \caption{Lifetime of object instances (log2 of number of cpu cycles) in Clang-11.}
  \label{fig:hist}
\end{figure}

\begin{figure}
  \includegraphics[width=0.5\textwidth]{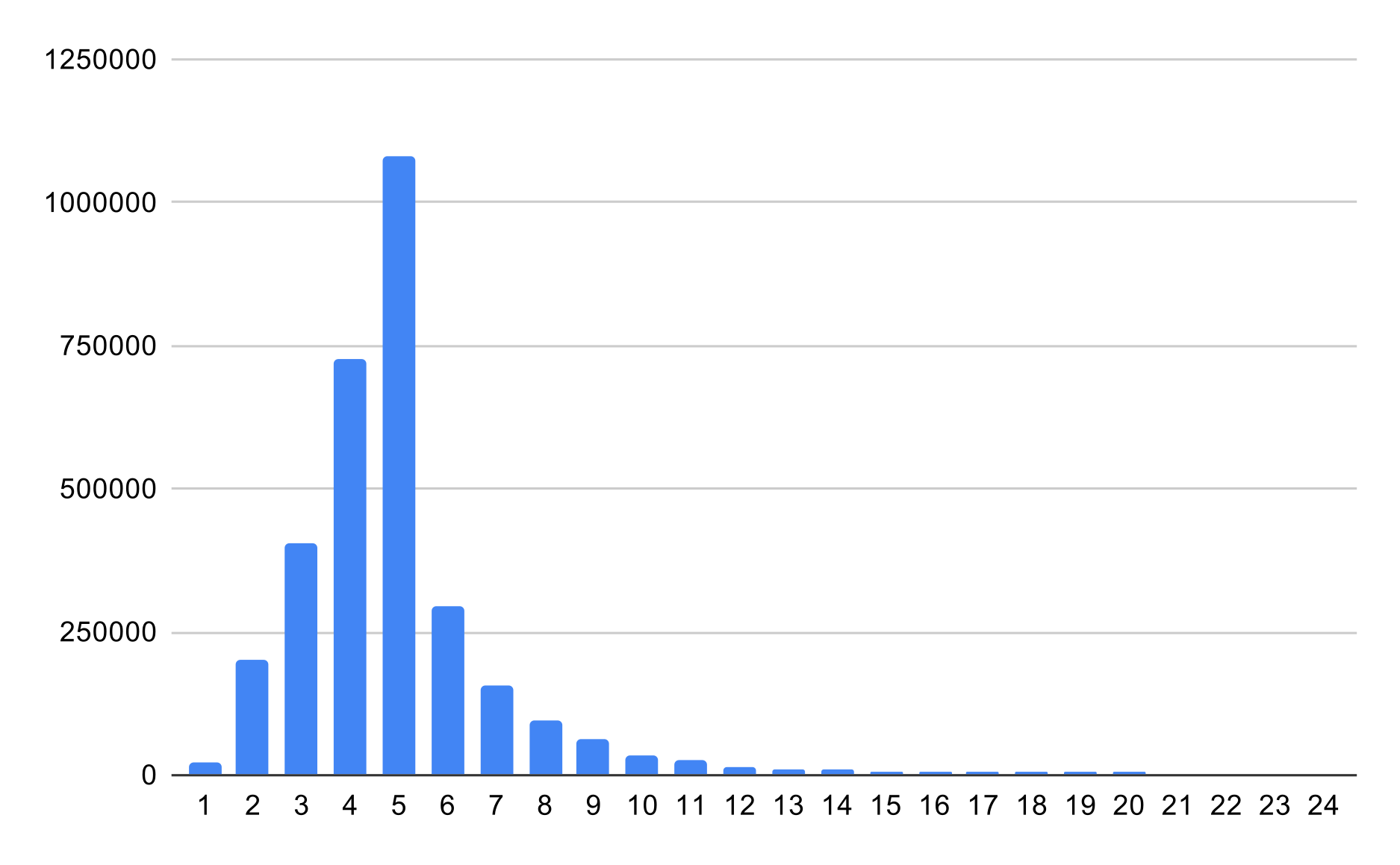}
  \caption{Number of shared\_ptr instances categorized by the max ref-count in libtorrent.}
  \label{fig:refcnt}
\end{figure}

Figure \ref{fig:hist2} presents a histogram that categorizes instances of string
and vector into different size buckets. Each size bucket represents the number
of elements when the destructor is called. This figure shows that most strings
and vectors are relatively small (below 64 elements), and that strings are more
frequently used than vectors.

Figure \ref{fig:hist} presents a histogram that categorizes the lifetime of
object instances from the C++ standard library (vector, string, shared\_ptr,
map, etc). The data for figures \ref{fig:hist2} and \ref{fig:hist} was collected
by tracing an execution of the Clang compiler on two large C input file
(oggenc.c and sqlite.c). The X-axis represents the number of cycles between the
construction and destruction of the object, in log2 representation.  The chart
presents both stack-allocated instances and heap-allocated instances.  It's
important to remember that stack-allocated instances of containers, such as
vector and string, can allocate memory dynamically.

The large number of instances that live for a short period of time present an
opportunity because perhaps some of these instances could be optimized away.
Some of the rules in the Warrior1 report tool were designed to detect these
short-lived objects and report them to the user.

Figure \ref{fig:refcnt} presents a histogram that categorizes instances of
shared\_ptr by the highest reference count that was observed for each instance
while tracing a program from the libtorrent library.

\begin{figure}
  \includegraphics[width=0.5\textwidth]{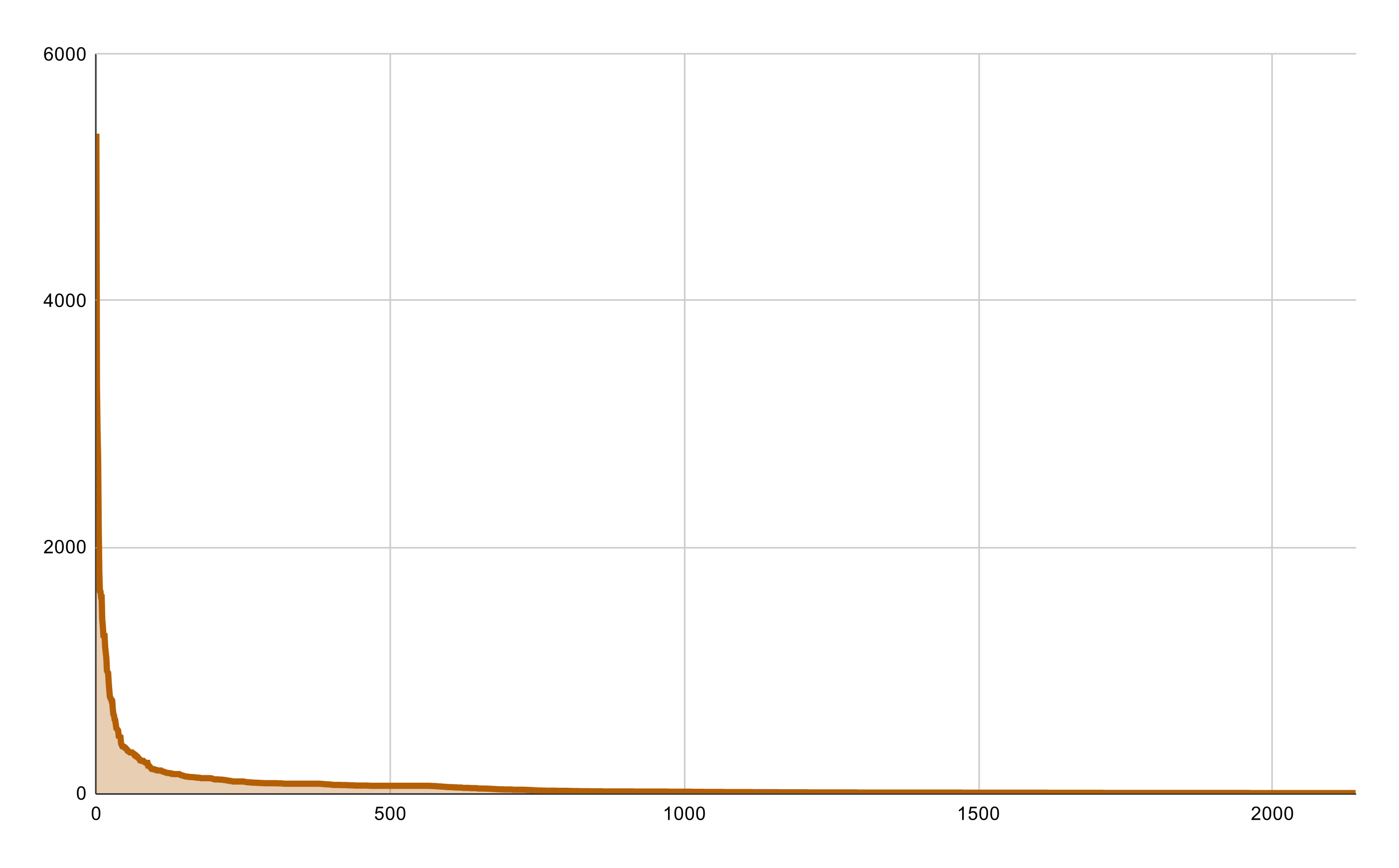}
  \caption{Number of string instances categorized by the number of string content repetition in Fish-shell.}
  \label{fig:strings}
\end{figure}

Figure \ref{fig:strings} presents a histogram that categorizes instances of
string by the number of times the string content appeared in previous strings.
When a string instance is passed by value it is copied and the content is
duplicated. This example was extracted from the Fish-shell program. In this
program, most string containers hold almost-unique string values, but some
non-empty string values are held by thousands of string instances. Perhaps some
of these values could be eliminated.

\section{Related work}

C++ is a high-level programming language that was designed\cite{cpp} as an extension to C.
Over the years C++ gained many features that interact in complex
ways.  The complexity of the language makes it difficult to write high-performance,
correct, maintainable and secure programs.  The C++ community overcame some of
the problems by introducing tools. Valgrind \cite{valgrind} and later the family of sanitizers,
and especially the address-sanitizer\cite{asan}, allowed users to catch memory-corruption
bugs. Better compiler analysis and diagnostics exposed some classes of bugs.
Clang-format and clang-tidy made the code more readable and maintainable.

High-level abstractions in C++ come with a performance cost. Compilers usually
can't recover the performance loss that are introduced with high-level
abstractions\cite{chandler}.  In this paper we highlighted the example of costly
calls to 'malloc', but there are also other examples of atomic operations or
just more instructions that are executed.

BOLT\cite{BOLT} removes some of the C++ abstraction overhead by reordering the
code according to the execution frequency. Moving the slow-path out of the way
helps to make better use of the instruction cache.

Another effective approach to improving the performance of C++ programs is making
memory allocations faster. JEMalloc\cite{jemalloc} is an efficient memory
allocator that is tuned for C++ programs.

Another approach to improving the performance of c++ programs is
the use of linters\cite{linter} that pattern match the AST and apply a set
of rules. Linters could catch patterns such as passing a vector by value instead
of by reference.  However, in practice, inefficient code patterns are complex
and often span different parts of the code and in different compilation units.
Moreover, some performance issues depend on the runtime behavior of the program.
For example, if a vector remains small we may prefer to replace it with a
non-allocating vector that has internal storage (such as Abseil's
InlinedVector, Folly's small\_vector or LLVM's SmallVector).

Perflint\cite{perflint} is similar to Warrior1. It is an analysis tool that
identifies suboptimal use patterns of C++ containers. Both are
implemented as instrumentation of the standard library and as diagnostics tool
that analyzes the recorded traces. There are a few differences between their
approach and ours. Perflint performs more of the analysis inside the host
program, while our approach relies more on an external diagnostics tool.
Perflint is focused on identifying the optimal data-structure, while our work
enables the detection of a wider set of patterns, such as the map count-operator
pair, or short-lived heap allocated objects.

Toddler\cite{toddler} detects opportunities for loop invariant code motion in
Java programs by inspecting traces of computation and memory access patterns.

\section{Examples}

This section brings a number of interesting examples from performance
anti-patterns that the Warrior1 tool discovered in popular open source prorgams.


The code below, in \textbf{LLVM} ($LiveVariables.cpp+93$), uses a
vector as a stack for a worklist. However, the tool found that the typical stack
size is relatively small. We replaced the code below with LLVM's SmallVector
data structure that contains on-stack inline storage and saved a large number of
memory allocations.  We found and fixed additional performance issues in LLVM:
D83849, D83788, D83797, D84620, D85538.

\begin{minipage}{\linewidth}
\lstset {language=C++}
\begin{lstlisting}[]
// Fixed in D83920
void LiveVariables::MarkVirtRegAliveInBlock(...) {
  std::vector<MachineBasicBlock*> WorkList;

  while (!WorkList.empty()) {
\end{lstlisting}
\end{minipage}

In the code below, from the \textbf{CMake} project ($cmListFileCache.cxx +372$),
a shared\_ptr is passed by value instead of by reference. Shared pointers should
be passed by reference because copying them is an expensive operation.

\begin{minipage}{\linewidth}
\lstset {language=C++}
\begin{lstlisting}[]
cmListFileBacktrace:: cmListFileBacktrace(
              std::shared_ptr<Entry const> top) {
...
\end{lstlisting}
\end{minipage}

In the code below, from the \textbf{Ninja-build} project ($manifest\_parser.cc +300$),
a new instance of EvalString is created on each
iteration of the loop. This data-structure contains a vector member that
allocates and deallocates memory on each iteration of the loop.  It is advised
to move the EvalString instance outside of the loop and clear the vectore before
entering the loop.

\begin{minipage}{\linewidth}
\lstset {language=C++}
\begin{lstlisting}[]
bool ManifestParser::ParseRule(string* err) {
...
while (lexer_.PeekToken(Lexer::INDENT)) {
  string key;
  EvalString value;
  if (!ParseLet(&key, &value, err))
    return false;
  ...
  }
\end{lstlisting}
\end{minipage}

In the code below, from the \textbf{Z3} project ($theory\_lra.cpp +1564$),
a map is searched twice. The first line searches the internal tree
data-structure of the map in search of the key, and after finding the key, the
second line searches the map again to return the key value.

\begin{minipage}{\linewidth}
\lstset {language=C++}
\begin{lstlisting}[]
lp::impq get_ivalue(theory_var v) const {
  ...
  auto t = get_tv(v);
  if (m_variable_values.count(t.index()) > 0)
      return m_variable_values[t.index()];
\end{lstlisting}
\end{minipage}

In the code below from the \textbf{Bitcoin} project ($bloom.cpp +247$), the
vector 'data' saves the temporary value from the variable hash. This code
pattern is inefficient because the vector allocates memory, and copies the
content of 'hash' before passing it to the non-mutating insert method (that
accepts a const vector). One possible fix would be to add a new method to the
class that accepts iterators or a data structure that is similar to LLVM's
ArrayRef or Abseil's Span.

\begin{minipage}{\linewidth}
\lstset {language=C++}
\begin{lstlisting}[]
void CBloomFilter::insert(const uint256& hash) {
  std::vector<unsigned char> data(hash.begin(), hash.end());
  insert(data);
}
\end{lstlisting}
\end{minipage}

In the code below, from the \textbf{ProtocolBuffers} project ($cpp\_message.cc +698$),
a map is constructed in some helper function. Later, each one of the keys and
values, which are strings that are individually allocated, are copied into
another map inside the AddMap method.

\begin{minipage}{\linewidth}
\lstset {language=C++}
\begin{lstlisting}[]
void MessageGenerator::
GenerateFieldAccessorDeclarations(...) {
...
  std::map<std::string, std::string> vars;
  SetCommonFieldVariables(field, &vars, options_);
  format.AddMap(vars);
...
\end{lstlisting}
\end{minipage}

In the code below, from the \textbf{Redex} project ($ControlFlow.cpp+1047$), is
written using a modern C++ style, where a loop is replaced by a for\_each and a
lambda method that constructs a temporary vector.  The tool detected this issue
because of the high number of memory allocations.  The code below was replaced
by a simple loop that counted the number of values and saved a large number of
memory allocations. We found and fixed additional performance issues in Redex:
D22828350, D22832242, D22832839.

\begin{minipage}{\linewidth}
\lstset {language=C++}
\begin{lstlisting}[]
// Fixed in 8cd1ee9
void ControlFlowGraph::sanity_check() const {
  ...
  auto num_succs =
  get_succ_edges_if(b, [](const Edge* e)
            { return e->type() != EDGE_GHOST; })
            .size();
\end{lstlisting}
\end{minipage}

In the code below, from the \textbf{ethereum/solidity} project ($Scanner.cpp +973$),
the struct CharStream is passed by value. This struct contains large strings
that are re-allocated when the struct is copied when passed by value.

\begin{minipage}{\linewidth}
\lstset {language=C++}
\begin{lstlisting}[]
void Scanner::reset(CharStream _source) {
\end{lstlisting}
\end{minipage}

In the code below, from the \textbf{libtorrent} project ($piece\_picker.cpp +350$), a vector
is returned by value. The vector is immediately destroyed and the memory is
deallocated. Returning the iterator over the internal member could be a more
efficient design. The average number of elements in the returned vector is 48,
so passing a small-vector as a parameter to the function is also a possible
design.

\begin{minipage}{\linewidth}
\lstset {language=C++}
\begin{lstlisting}[]
std::vector<piece_picker::downloading_piece>
piece_picker::get_download_queue() const {
  std::vector<downloading_piece> ret;
  for (auto const& c : m_downloads)
          ret.insert(ret.end(), c.begin(), c.end());
  return ret;
}
\end{lstlisting}
\end{minipage}

In the code below, from the \textbf{nlohmann-json} project($json.hpp +12734$), a
member of type output\_adapter\_t is returned by value. This type is an alias to
a shared\_ptr. This operation is expensive because passing a smart pointer by
value updates the reference count multiple times, and this is done with atomic
operations.

\begin{minipage}{\linewidth}
\lstset {language=C++}
\begin{lstlisting}[]
class output_adapter {
...
 operator output_adapter_t<CharType>()
 {
     return oa;
 }
...
    output_adapter_t<CharType> oa = nullptr;
};
\end{lstlisting}
\end{minipage}

In the code below, from the \textbf{Folly} project ($json.cpp +192$), the code
constructs two temporary string instances before appending the result into a
third string. These two string instances allocate memory if the size of the
string exceeds the internal storage.  A faster (and more readable code) would
just append the new line and the spaces in a loop.

\begin{minipage}{\linewidth}
\lstset {language=C++}
\begin{lstlisting}[]
void newline() const {
  if (indentLevel_) {
    out_ += to<std::string>('\n',
           std::string(*indentLevel_ * 2, ' '));
  }
}
\end{lstlisting}
\end{minipage}

In the code below, from the \textbf{Mcrouter}
project($FailoverErrorsSettingsBase.cpp+56$), the code passes a vector of
strings by value, instead of by reference. Passing the vector by value forces a
memory-allocating copy of the vector and of each one of the strings in the
vector.

\begin{minipage}{\linewidth}
\lstset {language=C++}
\begin{lstlisting}[]
void FailoverErrorsSettingsBase::
  List::init(std::vector<std::string> errors) {
\end{lstlisting}
\end{minipage}

In the code below, from the \textbf{PyTorch} project ($function\_schema\_parser.cpp+297$),
a function returns a union of two types, by value. The type FunctionSchema is a
complex struct that contains multiple strings, shared pointers and vectors of
structs that contain strings.  The function returns the struct by value and
copies the whole data structure.

\begin{minipage}{\linewidth}
\lstset {language=C++}
\begin{lstlisting}[]
C10_EXPORT either<OperatorName, FunctionSchema> parseSchemaOrName(...) {
  return SchemaParser(schemaOrName)
          .parseDeclarations().at(0);
}
...
auto parsed = parseSchemaOrName(schema);
\end{lstlisting}
\end{minipage}

The code below, from the \textbf{Inkscape} project($color-item.cpp +464$), was
selected because it presents four different inefficiencies that were discovered
by the tool. First, the vector $entries$ grows one element at a time, but the
size of the vector could be reserved when the vector is initialized, because the
loop trip count is known. Second, the size of the vector is typically 3-4
elements, which means that we could use a non-allocating small-vector. Third,
the string $str$ in the for-each loop is copied by value instead of accessed by
reference.  Finally, elements are inserted into the vector using the
$push\_back$ method, instead of the $emplace\_back$ method. This is inefficient
because the struct $entry$ is copied right after it is constructed.  However,
the biggest issue that Warrior1 found in Inkscape are in the method
Bezier::roots (not shown here) that returns a vector of Bezier coordinates. The
methods boundsExact and feed\_curve\_to\_cairo call this API very frequently,
allocating memory on each invocation of the function. A more efficient API would
allow passing a pre-allocated storage by reference. In

\begin{minipage}{\linewidth}
\lstset {language=C++}
\begin{lstlisting}[]
std::vector<Gtk::TargetEntry> entries;

for (auto str : listing) {
  auto target = str.c_str();
  guint flags = 0;
  if (mimeToInt.find(str) == mimeToInt.end()) {
  ...
  }
  auto info = ... 
  Gtk::TargetEntry entry(target, flags, info);
  entries.push_back(entry);
}

preview->drag_source_set(entries, Gdk::BUTTON1_MASK, Gdk::DragAction(...) );
\end{lstlisting}
\end{minipage}

In the code below, from the \textbf{GCC/libstdc++}
project($regex.h+2982$), a function uses a standard vector to store the input and transform it.
The typical input size is very small and could fit in a small vector and avoid a
malloc. This code is found inside the C++ standard library itself.

\begin{minipage}{\linewidth}
\lstset {language=C++}
\begin{lstlisting}[]
/**
 * @brief Gets a sort key for a character ...
 */
template<typename _Fwd_iter> string_type
  transform_primary(...) const {
    ...
    std::vector<char_type> __s(__first, __last);
    __fctyp.tolower(..);
    return this->transform(...);
  }
\end{lstlisting}
\end{minipage}

\section{Challenges}

Warrior1 is effective in detecting inefficient code sequences, but there are a
few challenges that prevent the tool from being helpful in some environments.

One challenge is that in many cases the inefficiency in the program is built
into some widely used API, and this makes it difficult to fix the program. For
example, in LLVM's table-gen tool a commonly used class declares a virtual
method with the name "getName" that returns a string. This API is inefficient
because returning a string forces memory allocation. The method is overridden by
dozens of other methods that implement the same API. The tool detects a huge
number of short-lived strings that allocate memory, but it is very difficult to
fix the issues because of the large number of overriding methods and call sites.

Another limitation of the tool is that some large-scale C++ projects often
implement their own data structures instead of using the standard library. One
of the reasons is that older C++ projects started before the standard library
was mature and reliable. Another reason is that many projects implement their
own high-performance data structures (such as Folly, Absail). These data
structures can't be analyzed by the tool without adding the logging methods and
special diagnostics code in the reporting tool.

Another challenge in applying the tool is that some software projects are
already well optimized. For example, the inefficiencies that we found in
Webkit's JavaScript engine were not significant because engineers spent a lot of
time optimizing the code. There are not that many opportunities to remove
inefficient code in projects that are regularly inspected using a profiler by
many people.

\section{Acknowledgements}

We'd like to thank Evgeny Fiksman and Maged Michael for their help in running
the tool on Facebook workloads and in detecting and fixing issues in production
services.

\section{Conclusion}
The C++ community is very familiar with the problem of inefficient code that's
caused by the design of the language and the implementation of complex
abstractions.  This paper presents Warrior1, a tool for detecting inefficient
code patterns in C++. Warrior1 was designed to detect the long tail of small
performance issues that are difficult to detect with a profiler, and that are
the result of inefficient use of C++ libraries.  We instrument the standard
library to log operations and stack traces in a compressed binary format.  We
show that the tool is effective in detecting inefficient code patterns in a
number of open source software projects. We fixed some of the issues that the
tool detected in open source programs.  We demonstrate that C++ containers are
frequency misused in real programs (including in already relatively
well-optimized ones, written by well-regarded C++ programmers, like LLVM).  The
techniques that we presented in this paper could be applicable to other
programming languages, such as Rust, Swift and Java.

\bibliographystyle{plain}
\bibliography{paper.bib}

\end{document}